\documentclass[aps,11pt]{revtex4}
\usepackage[dvips]{graphicx}

\usepackage{epsfig}
\usepackage{amsmath}
\usepackage{bm}

\begin{document}

\preprint{}

\title{Can we distinguish between $h^{SM}$ and $h^0$ in 
split supersymmetry?}
\author{Marco Aurelio D\'\i az}
\author{Pavel Fileviez P\'erez}
\affiliation{Pontificia Universidad Cat\'olica de Chile \\
Facultad de F{\'\i}sica, Casilla 306 \\
Santiago 22, Chile.}
\begin{abstract}
We investigate the possibility to distinguish between the Standard
Model Higgs boson and the lightest Higgs boson in Split Supersymmetry. 
We point out that the best way to distinguish between 
these two Higgs bosons is through the decay into two photons. 
It is shown that there are large differences of several percent between the 
predictions for $\Gamma(h\rightarrow\gamma\gamma)$ in the two models, 
making possible the discrimination at future photon-photon colliders. 
Once the charginos are discovered at the next generation of collider experiments, 
the well defined predictions for the Higgs decay into two photons 
will become a cross check to identify the light Higgs boson 
in Split Supersymmetry.                    
\end{abstract}
\pacs{}
\maketitle

\section{Introduction}

Despite the great success of the Standard Model (SM), the mechanism 
for electroweak symmetry breaking remains to be tested in experiments. 
There are many reasons to believe there is physics beyond the SM. In 
particular, the Minimal Supersymmetric Standard Model (MSSM) is one of 
the best motived theories where it is possible to describe the cold 
dark matter in the Universe and where the unification of the gauge 
couplings is achieved. In low energy supersymmetry it is assumed that 
the SUSY breaking scale is at TeV following the naturalness criterion. 

Recently, Arkani-Hamed and Dimopoulos \cite{Arkani}, have noticed that 
gauge coupling unification can be achieved in a supersymmetric model 
where all scalars, except for one Higgs doublet, are very heavy. 
Most of the unpleasant aspects of low energy
supersymmetry, such as excessive flavour and CP violation, very fast 
dimension 5 proton decay, and tight constraints on the Higgs mass, are 
eliminated. At the same time, there is a candidate to describe the 
cold dark matter in the Universe. Several phenomenological studies 
in this scenario, now called split supersymmetry, have been performed
\cite{Split1,Split2,Split3,Split4,Split5,Split6,Split7,
Split8,Split9,Split10,Split11,Split12,Split13,Split14,Split15,
Split16,Split17}.

It is expected that the Higgs boson will be discovered in the next 
generation of collider experiments. Therefore, one of the main issues 
in Higgs boson physics is the identification of observable useful to 
distinguish between the SM Higgs $h^{SM}$ and the lightest Higgs in 
possible extensions of the SM. There are several studies devoted to 
this very important issue in the context of the MSSM
\cite{Georgi,Wilzeck,Cahn,Hulth,Ellis,Shifman,Djouadi,Akeroyd:2003bt,
Akeroyd:2003xi,Babu,Hollik,Carena}.

As we mention before we hope at future collider experiments 
we will discover at least a neutral particle with spin zero, 
the Higgs boson, and the lightest supersymmetric particles, 
the neutralinos and charginos. Since in Split 
supersymmetry the interactions of the 
lightest Higgs boson with the SM fermions are the same as the 
interactions of the SM Higgs, through the decays into two 
SM fermions it is not possible to make the identification of the Higgs. 
Therefore we have to use the decays at one loop, 
where the effect of virtual particles present in Split SUSY is relevant. 
In this Letter we investigate this issue and we argue that the 
way to address it is studying the decays of the Higgs boson 
into two photons. We show several numerical examples where it is 
possible to appreciate large differences between the decay rates in the 
SM and in Split supersymmetry.        

\section{Distinguishing between $h^{SM}$ and $h^0$ in Split SUSY}

In order to identify the Higgs boson at future 
colliders the predictions of all its decay rates have to match with 
their measurements. This is the case of SM Higgs boson 
$h^{SM}$ or the lightest Higgs boson $h^0$ of 
the supersymmetric extension of the standard model. 

It is particularly difficult to distinguish between the SM Higgs boson and 
the MSSM lightest Higgs boson in the decoupling limit \cite{Haber:2002ey}.
This is precisely what happens in Split Supersymmetry, where all
scalars are very heavy except for one Higgs doublet and for the neutralinos 
and charginos which are all light. Therefore, in split supersymmetry we expect 
that at future colliders the lightest Higgs boson will be discovered 
together with charginos and neutralinos. However, their discovery is not 
enough to claim that the observed Higgs corresponds to the SPLIT SUSY 
light Higgs, since we have to measure its couplings with a good precision.

In the context of the MSSM there are studies about 
the possibility to distinguish between different 
Higgses through the couplings $Hgg$ \cite{Georgi}, $H Z \gamma$ 
\cite{Wilzeck,Cahn,Hulth} and $H \gamma \gamma$ 
\cite{Ellis,Shifman,Djouadi,Akeroyd:2003bt,Akeroyd:2003xi}. Also the 
quantity  $B(H \to b \bar{b})/ B(H \to \tau^+ \tau^-)$ has been studied 
extensively \cite{Babu,Hollik,Carena}. 

In this Letter we study the possibility to identify the light Higgs 
in split supersymmetry \cite{Arkani}. Since in this case all sfermions 
and Higgs bosons, except for the lightest 
one, are very heavy it is not possible to use the quantity 
$B(H \to b \bar{b})/ B(H \to \tau^+\tau^-)$ nor the coupling $Hgg$. 
The $Hb\bar{b}$ and $H\tau\tau$ couplings are equal at tree level 
in the SM and in SPLIT SUSY. There are not differences at 
one-loop, since all squarks or sleptons in the 
loops are very heavy. In the case of $Hgg$ coupling, this is a
one-loop induced coupling with quark contributions being common to both 
the SM and SPLIT-SUSY, and with the difference being the squark 
contributions, which are negligible.
The two other possibilities are to use the couplings 
$H Z \gamma$ and $H \gamma \gamma$. However, it is not possible to use 
the first one since we know that the $B(h \to Z \gamma)$ is very small 
and it will be very difficult to determine it with a very good precision 
\cite{Dubinin:2003kb}. Therefore we concentrate on the possibility to 
use the coupling $H \gamma \gamma$ for our study. The decay 
$h\rightarrow\gamma\gamma$ can be observed at the LHC but with an error 
larger than $10\%$ \cite{Carena:2002es,Bern:2002jx,Group:2004hn}. 
At electron-positron colliders there is the additional possibility of 
photon-photon collisions, where the rate
$\gamma\gamma\rightarrow h\rightarrow b\bar b$ can be measured with a 
$2\%$ precision 
\cite{Asner:2002aa,Asner:2001vh,Soldner-Rembold:2000pb,Gunion:2003fd}. 
This combined with a measurement of $B(h\rightarrow b\bar b)$ with a 
$2.4\%$ \cite{Group:2004hn}, gives a determination of 
$B(h\rightarrow\gamma\gamma)$ with a $\sim2\%$ error.

The decay rate $\Gamma(h \to \gamma \ \gamma)$ is 
given by \cite{HHG}:
\begin{equation}
\Gamma(h \to \gamma \ \gamma)= \frac{\alpha^2 g^2}{1024 \pi^3}
\frac{m_h^3}{m_W^2}\left|\sum_{i} A_i \right|^2
\end{equation}
where $\alpha$ is the fine structure constant, $g$ is the $SU(2)$ gauge
coupling constant, $m_W$ is the $W$ boson mass, and $m_h$ is the Higgs
boson mass. There is an amplitude $A_i$ for each charged particle
inside the loop, and depends on a loop function $F$ whose form varies
according to the spin of the particle in the loop.

In Split Supersymmetry the relevant contributions are:
\begin{eqnarray}
A_W & = & C_{W} F_1(\tau_W) \\
A_f & = & N_C^f \ Q_f^2 \ C_{f} \ F_{1/2}(\tau_f) \\ 
A_{\tilde{\chi}^{\pm}} & = & C_{\tilde{\chi}^{\pm}} 
\frac{M_W}{M_{\tilde{\chi}^{\pm}}} 
F_{1/2}(\tau_{\tilde{\chi}^{\pm}}) 
\end{eqnarray}
where $A_W$, $A_f$, $A_{\tilde{\chi}^{\pm}}$ are the amplitudes for the 
contributions with $W$ bosons, fermions, and charginos inside the loop, 
respectively. The parameter $\tau_i = 4 m_i^2/ m^2_h$, where $m_i$ is 
the mass of the particle inside the loop, and $m_h$ is the Higgs mass. 
$N_C =3$ for quarks and squarks, while $N_C =1$ for leptons and sleptons, 
and $Q_f$ is the electric charge of the fermion $f$. 
The couplings of the lightest Higgs to the internal particles 
are given by:
\begin{equation}
C_{f=u,c,t} = 1,  \ \ C_{f=d,s,b,e,\mu,\tau} = 1, \ \ C_W = 1, 
\end{equation}
\begin{equation} 
C_{\tilde{\chi}_i^{\pm}} =  2 (S_{ii} \sin \beta + Q_{ii} \cos \beta)
\end{equation}
with $S_{ij}= U_{i1} V_{j2} / \sqrt{2}$ and 
$Q_{ij}= U_{i2} V_{j1} / \sqrt{2}$. The matrices $U$ and $V$ are 
the matrices which diagonalize the chargino mass matrix. 
The loop functions $F_0$, $F_{1/2}$ and $F_1$ can be found in \cite{HHG}.

In Split supersymmetry, the Higgs couplings to fermions 
and $W$ bosons are SM-like, giving contributions which are equal 
in both models. Therefore, the chargino contribution will 
determinate the difference between the decays into two photons 
in the SM and split SUSY.
%
\section{Results}

In order to quantify the difference between the decay rate of the SM 
Higgs and the lightest Higgs into two photons in split supersymmetry 
we define the following quantity:

\begin{equation}
\delta = \frac{\Gamma^{Split}(h^0 \to \gamma \gamma)- \Gamma(h^{SM} \to \gamma \gamma)}{\Gamma(h^{SM} \to \gamma \gamma)}
\end{equation} 

We calculate this quantity using the above formulas for different values 
of the relevant parameters, which are the Higgs mass, the higgsino mass 
parameter $\mu$, the $SU(2)$ gaugino mass $M_2$, and $\tan\beta$. We take
$m_h=120$ GeV and consider it as an independent parameter.
Notice that in the case of split supersymmetry, since we accept the 
fine-tuning and we integrate out all superheavy scalars, the Higgs mass 
basically does not change when we vary the rest of the parameters. The 
loops corrections with charginos and neutralinos are not very important, 
see for example \cite{Split8}.   

\begin{figure}
\centerline{\protect\hbox{\epsfig{file=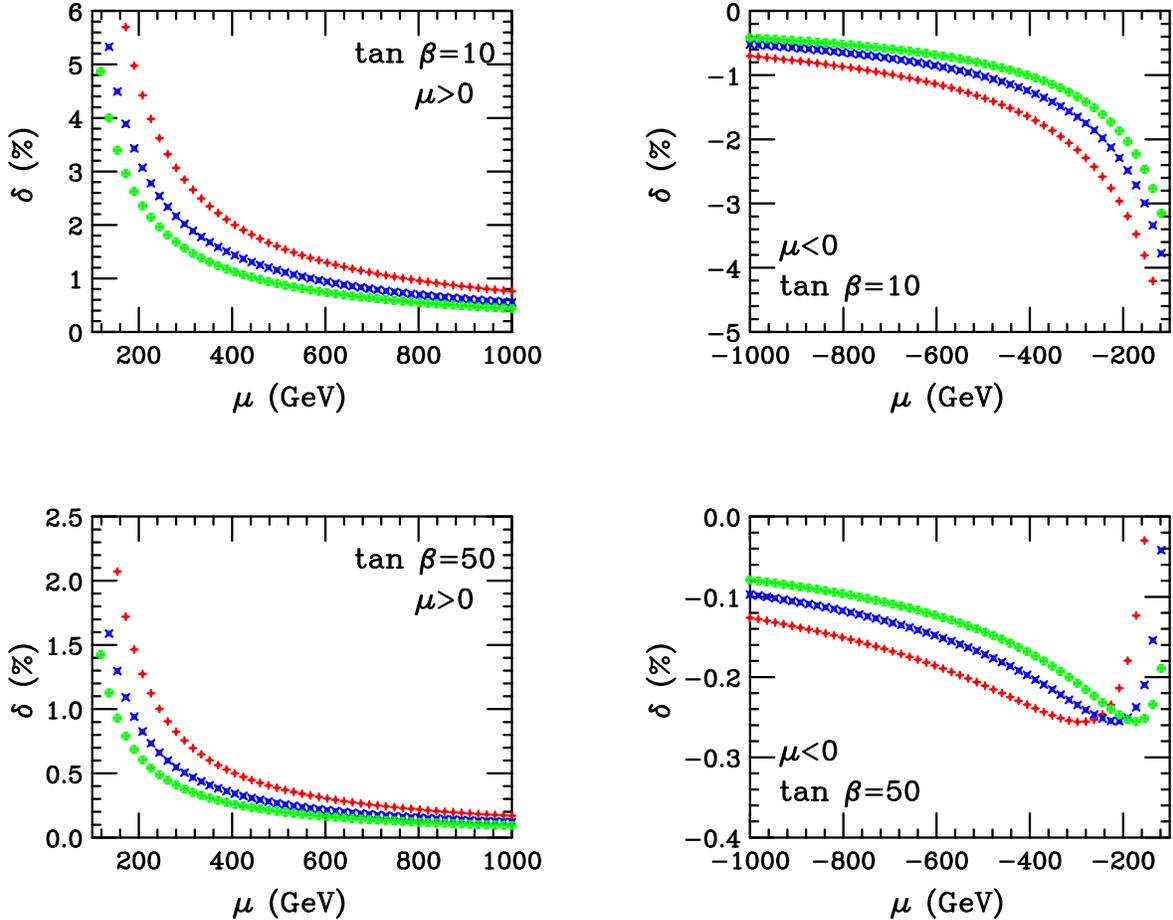,width=0.75\textwidth,angle=90}}}
\caption{\it Relative difference $\delta$ between 
$\Gamma(h\rightarrow\gamma\gamma)$ in the SM and in split supersymmetry
as a function of $\mu$ for different values of $\tan\beta$ and $M_2$.
Red crosses correspond to $M_2=150$ GeV, blue x's to $M_2=200$ GeV, and
green circles to $M_2=250$ GeV. We consider $m_h=120$ GeV.
}
\label{dmu}
\end{figure} 

In Figure \ref{dmu} we see how this quantity changes as a function of
the parameter $\mu$, for given values of $M_2$ and $\tan \beta$. 
Note that it is possible to achieve differences up to $6\%$ when the 
chargino contribution is large. As this difference is achieved, the 
branching ratio of the decay into two photons is of the order of 
$10^{-3}$. We impose the experimental bound $m_{\tilde\chi_1^+}>103$ GeV,
which limits the curves at low values of $\mu$. As the magnitude of
the parameter $\mu$ increases the chargino masses increase also, the
heavier faster than the lightest. In general, larger chargino masses
produce smaller contributions to $h\rightarrow\gamma\gamma$ and thus
the parameter $\delta$ decreases. For a similar reason, curves defined
by larger values of the gaugino mass $M_2$ have smaller
$\delta$. These curves are defined by $M_2=150$ GeV (red crosses), 
$M_2=200$ GeV (blue x's), and $M_2=250$ GeV (green circles). Two
different values of $\tan\beta$ are considered in this figure, 
$\tan\beta=10$ and 50 with the smaller value giving larger $\delta$.
Another interesting point to notice in this figure is the correlation
between the sign of $\mu$ (actually, the sign of $\mu M_2$ since we work 
with $M_2>0$) and the sign of $\delta$.

\begin{figure}
\centerline{\protect\hbox{\epsfig{file=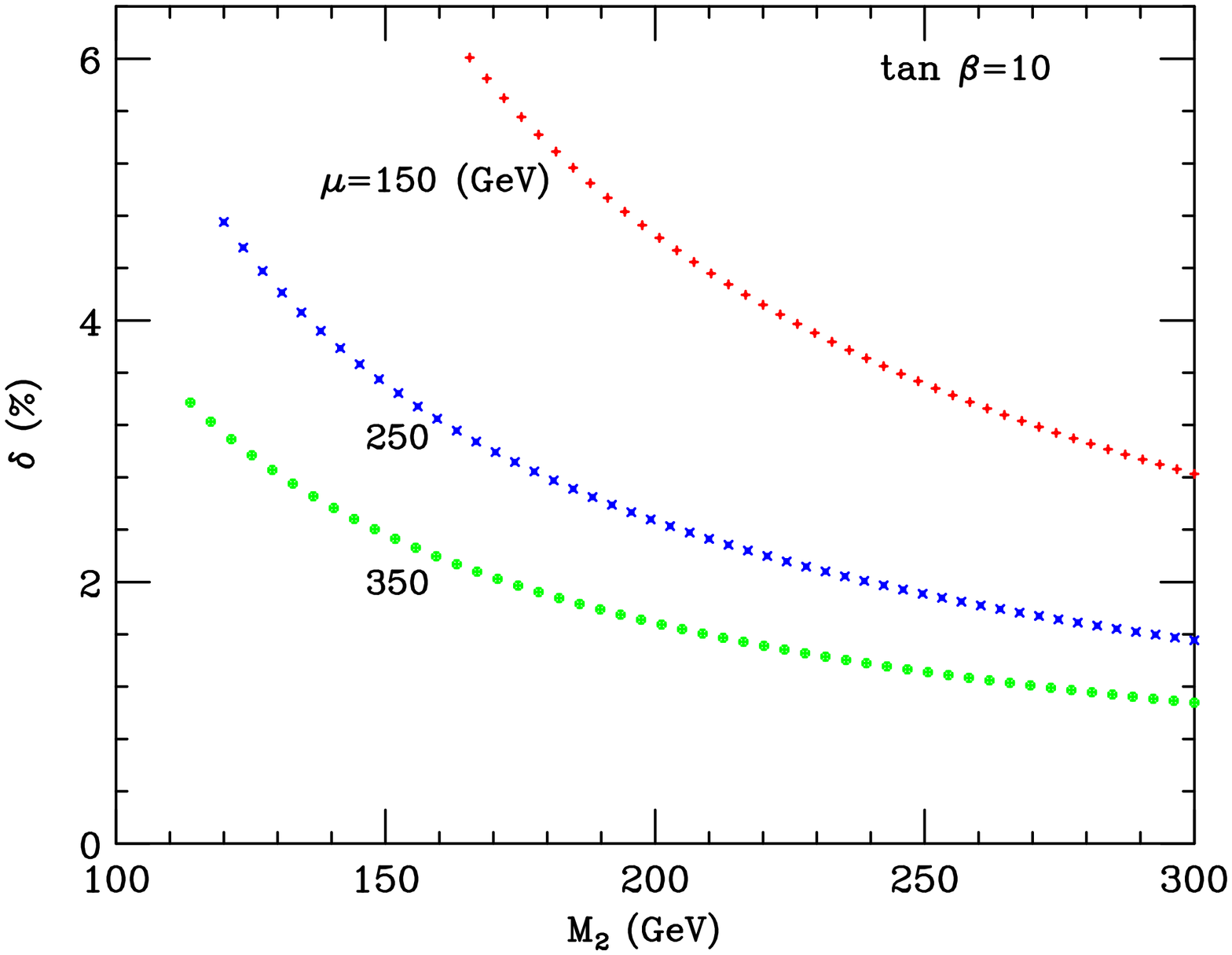,width=0.5\textwidth,angle=90}}}
\caption{\it Relative difference $\delta$ between 
$\Gamma(h\rightarrow\gamma\gamma)$ in the SM and in split supersymmetry
as a function of $M_2$, for $\tan\beta=10$ and different values of
$\mu$. We consider $m_h=120$ GeV.
}
\label{dm2}
\end{figure} 

In Fig.~\ref{dm2} we show the variation of $\delta$ as a function of 
$M_2$ for $\tan\beta=10$ and $\mu=150$, 250, and 350 GeV. It is clear 
that when $M_2$ is increased our quantity decreases since the chargino 
contributions are less important, due to their larger mass. The three
curves are limited at low $M_2$ by the experimental lower bound on the
chargino mass described before. At the other extreme ($M_2=300$ GeV)
the light chargino mass is given by 133, 213, and 262 GeV for 
$\mu=150$, 250, and 350 GeV, respectively. Despite these large chargino
masses, $\delta$ remains above 1\% in the whole parameter space shown
in the figure.

\begin{figure}
\centerline{\protect\hbox{\epsfig{file=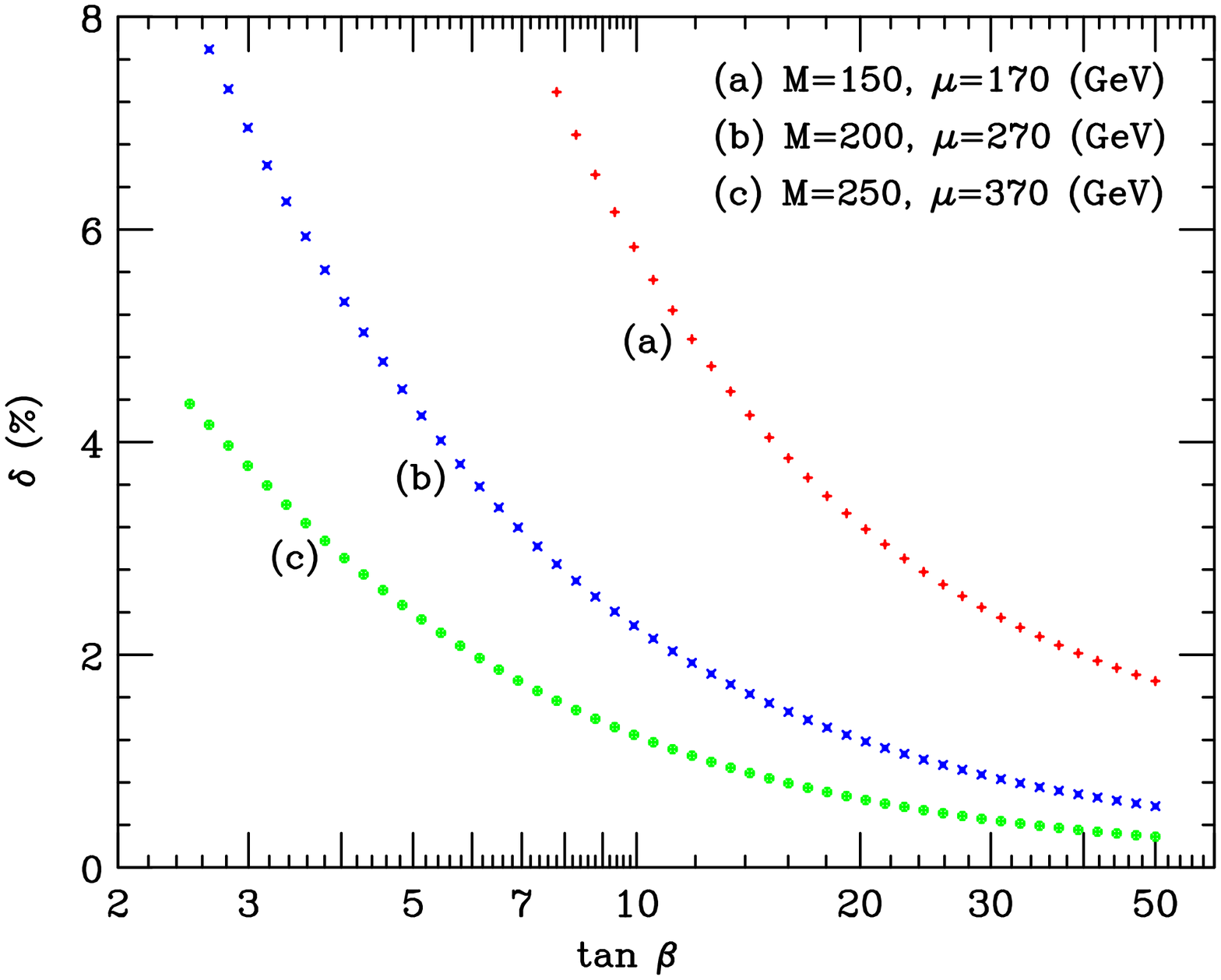,width=0.5\textwidth,angle=90}}}
\caption{\it Relative difference $\delta$ between 
$\Gamma(h\rightarrow\gamma\gamma)$ in the SM and in split supersymmetry
as a function of $\tan\beta$ for different values of $\mu$ and $M_2$.
We consider $m_h=120$ GeV.
}
\label{dtb}
\end{figure} 

In Fig.~\ref{dtb} we have the dependence of $\delta$ on $\tan\beta$,
and plot three curves with increasing values of $\mu$ and $M_2$, as
indicated in the figure. In all cases $\delta$ decreases with
$\tan\beta$ from several percent at low values to less than 1-2\% at
large values.

The decay rate for $h\rightarrow\gamma\gamma$ in the MSSM with the Higgs
sector in the decoupling limit was studied in \cite{Djouadi,Carena}. 
It was found that contributions from the stop sector are larger when 
the left-right mixing is large, and that contributions from charginos
decrease with $\tan\beta$. At small $\tan\beta$ chargino and 
stop contributions have opposite signs and a cancellation could occurs. 
Of course, this cancellation does not happens in Split Supersymmetry, 
obtaining larger decay rates $\Gamma(h\rightarrow\gamma\gamma)$. 
In general, smaller decay rates are obtained at large values of 
$\tan\beta$: chargino contributions are small in both the MSSM and 
Split SUSY, with the stop contribution adding to the $W$ contribution in 
the MSSM. Larger decay rates are found in the MSSM at 
high $\tan\beta$ when the sbottom loops are also important. 
We hope that the results presented in this section will be useful 
to understand the possibility to distinguish between the 
SM Higgs boson and the lightest Higgs boson in Split SUSY.

\section{Summary}

In Split Supersymmetry the light Higgs boson couplings
to SM particles are identical to the couplings of the SM Higgs boson. 
We point out that a way to distinguish between them
is through the decay into two photons. We show several numerical examples 
where we appreciate large differences of several percent between the 
predictions for $\Gamma(h\rightarrow\gamma\gamma)$ in the two models, 
making possible the discrimination at future photon-photon colliders. 
Once the Higgs boson and the charginos are discovered at the next 
generation of collider experiments, the well defined predictions 
for the Higgs decay into two photons will constitute an important 
cross check to identify the light Higgs boson in Split SUSY.

\begin{acknowledgments}
The work of P.F.P. and M.A.D. have been supported by 
CONICYT/FONDECYT under contract $N^{\underline 0} \ 3050068$ and
$N^{\underline 0} \ 1040384$, respectively.
\end{acknowledgments}



\begin{thebibliography}{99}

\bibitem{Arkani}
N.~Arkani-Hamed and S.~Dimopoulos,
arXiv:hep-th/0405159.

\bibitem{Split1}
G.~F.~Giudice and A.~Romanino,
Nucl.\ Phys.\ B {\bf 699} (2004) 65
[arXiv:hep-ph/0406088].

\bibitem{Split2}
N.~Arkani-Hamed, S.~Dimopoulos, G.~F.~Giudice and A.~Romanino,
arXiv:hep-ph/0409232.

\bibitem{Split3}
A.~Arvanitaki, C.~Davis, P.~W.~Graham and J.~G.~Wacker,
arXiv:hep-ph/0406034.

\bibitem{Split4}
A.~Pierce,
Phys.\ Rev.\ D {\bf 70} (2004) 075006
[arXiv:hep-ph/0406144].

\bibitem{Split5}
S.~h.~Zhu,
Phys.\ Lett.\ B {\bf 604} (2004) 207
[arXiv:hep-ph/0407072].

\bibitem{Split6}
B.~Mukhopadhyaya and S.~SenGupta,
arXiv:hep-th/0407225.

\bibitem{Split7}
R.~Mahbubani,
arXiv:hep-ph/0408096

\bibitem{Split8}
M.~Binger,
arXiv:hep-ph/0408240.


\bibitem{Split9}
J.~L.~Hewett, B.~Lillie, M.~Masip and T.~G.~Rizzo,
JHEP {\bf 0409} (2004) 070
[arXiv:hep-ph/0408248].


\bibitem{Split10}
L.~Anchordoqui, H.~Goldberg and C.~Nunez,
arXiv:hep-ph/0408284.

\bibitem{Split11}
I.~Antoniadis and S.~Dimopoulos,
arXiv:hep-th/0411032.

\bibitem{Split12}
B.~Bajc and G.~Senjanovic,
arXiv:hep-ph/0411193.

\bibitem{Split13}
B.~Kors and P.~Nath,
arXiv:hep-th/0411201.

\bibitem{Split14}
E.~J.~Chun and S.~C.~Park,
arXiv:hep-ph/0410242.

\bibitem{Split15}
K.~Cheung and W.~Y.~Keung,
arXiv:hep-ph/0408335.

\bibitem{Split16}
W.~Kilian, T.~Plehn, P.~Richardson and E.~Schmidt,
arXiv:hep-ph/0408088.

\bibitem{Split17}
A.~Masiero, S.~Profumo and P.~Ullio,
arXiv:hep-ph/0412058.


\bibitem{Georgi}
H.~M.~Georgi, S.~L.~Glashow, M.~E.~Machacek and D.~V.~Nanopoulos,
Phys.\ Rev.\ Lett.\  {\bf 40} (1978) 692.

\bibitem{Wilzeck}
F.~Wilczek,
Phys.\ Rev.\ Lett.\  {\bf 39} (1977) 1304.

\bibitem{Cahn}
R.~N.~Cahn, M.~S.~Chanowitz and N.~Fleishon,
Phys.\ Lett.\ B {\bf 82} (1979) 113.

\bibitem{Hulth}
L.~Bergstrom and G.~Hulth,
Nucl.\ Phys.\ B {\bf 259} (1985) 137
[Erratum-ibid.\ B {\bf 276} (1986) 744].

\bibitem{Ellis}
J.~R.~Ellis, M.~K.~Gaillard and D.~V.~Nanopoulos,
Nucl.\ Phys.\ B {\bf 106} (1976) 292.

\bibitem{Shifman}
M.~A.~Shifman, A.~I.~Vainshtein, M.~B.~Voloshin and V.~I.~Zakharov,
Sov.\ J.\ Nucl.\ Phys.\  {\bf 30} (1979) 711
[Yad.\ Fiz.\  {\bf 30} (1979) 1368].

\bibitem{Djouadi}
A.~Djouadi, V.~Driesen, W.~Hollik and J.~I.~Illana,
Eur.\ Phys.\ J.\ C {\bf 1} (1998) 149
[arXiv:hep-ph/9612362].

\bibitem{Akeroyd:2003bt}
A.~G.~Akeroyd and M.~A.~Diaz,
Phys.\ Rev.\ D {\bf 67} (2003) 095007
[arXiv:hep-ph/0301203].

\bibitem{Akeroyd:2003xi}
A.~G.~Akeroyd, M.~A.~Diaz and F.~J.~Pacheco,
Phys.\ Rev.\ D {\bf 70} (2004) 075002
[arXiv:hep-ph/0312231].

\bibitem{Babu}
K.~S.~Babu and C.~F.~Kolda,
Phys.\ Lett.\ B {\bf 451} (1999) 77
[arXiv:hep-ph/9811308].

\bibitem{Hollik}
J.~Guasch, W.~Hollik and S.~Penaranda,
Phys.\ Lett.\ B {\bf 515} (2001) 367
[arXiv:hep-ph/0106027].

\bibitem{Carena}
M.~Carena, H.~E.~Haber, H.~E.~Logan and S.~Mrenna,
Phys.\ Rev.\ D {\bf 65} (2002) 055005
[Erratum-ibid.\ D {\bf 65} (2002) 099902]
[arXiv:hep-ph/0106116].

\bibitem{Haber:2002ey}
H.~E.~Haber,
Nucl.\ Phys.\ Proc.\ Suppl.\  {\bf 116} (2003) 291
[arXiv:hep-ph/0212010].

\bibitem{Dubinin:2003kb}
M.~Dubinin, H.~J.~Schreiber and A.~Vologdin,
Eur.\ Phys.\ J.\ C {\bf 30} (2003) 337
[arXiv:hep-ph/0302250].

\bibitem{Carena:2002es}
M.~Carena and H.~E.~Haber,
Prog.\ Part.\ Nucl.\ Phys.\  {\bf 50} (2003) 63
[arXiv:hep-ph/0208209].

\bibitem{Bern:2002jx}
Z.~Bern, L.~J.~Dixon and C.~Schmidt,
Phys.\ Rev.\ D {\bf 66} (2002) 074018
[arXiv:hep-ph/0206194].

\bibitem{Group:2004hn}
L.~S.~Group {\it et al.},
arXiv:hep-ph/0410364.

\bibitem{Asner:2002aa}
D.~Asner, B.~Grzadkowski, J.~F.~Gunion, H.~E.~Logan, V.~Martin, 
M.~Schmitt and M.~M.~Velasco,
arXiv:hep-ph/0208219.

\bibitem{Asner:2001vh}
D.~Asner {\it et al.},
Eur.\ Phys.\ J.\ C {\bf 28} (2003) 27
[arXiv:hep-ex/0111056].

\bibitem{Soldner-Rembold:2000pb}
S.~Soldner-Rembold and G.~Jikia,
Nucl.\ Instrum.\ Meth.\ A {\bf 472} (2001) 133
[arXiv:hep-ex/0101056].

\bibitem{Gunion:2003fd}
J.~F.~Gunion, H.~E.~Haber and R.~Van Kooten,
arXiv:hep-ph/0301023.

\bibitem{HHG}
J. F. Gunion et al, Higgs Hunter's Guide.

\end{thebibliography}
\end{document}